\documentclass[10pt,letterpaper]{article}
\usepackage{opex3}
\usepackage{amssymb}
\usepackage{amsmath}
\usepackage{appendix}
\usepackage{cite}
\begin{document}

\title{Efficient noiseless linear amplification for light fields with larger amplitudes}

	\author{Jinwoo Park$^1$, Jaewoo Joo$^2$, Alessandro Zavatta$^{3, 4}$, Marco Bellini$^{3, 4}$, and Hyunseok Jeong$^{1,*}$}
	\address{
	$^1$Center for Macroscopic Quantum Control, Department of Physics and Astronomy, Seoul National University, Seoul 151-742, Republic of Korea\\
	$^2$Advanced Technology Institute and Department of Physics, University of Surrey, Guildford, GU2 7XH, United Kingdom\\
	$^3$Istituto Nazionale di Ottica, INO-CNR, Largo Enrico Fermi, 6, I-50125 Firenze, Italy\\
	$^4$Department of Physics and LENS, University of Firenze, 50019 Sesto Fiorentino, Firenze, Italy}
	\email{$^*$h.jeong37@gmail.com}


\begin{abstract}
We suggest and investigate a scheme for non-deterministic noiseless linear amplification of coherent states using successive photon addition, $(\hat a^{\dag})^2$, 
where $\hat a^\dag$ is the photon creation operator.
We compare it with a previous proposal using the photon addition-then-subtraction, $\hat a \hat a^\dag$, where $\hat a$ is the photon annihilation operator, that works as an appropriate amplifier only for weak light fields.
We show that when the amplitude of a coherent state is $|\alpha| \gtrsim 0.91$,
the $(\hat a^{\dag})^2$ operation serves as a  more efficient amplifier compared to the $\hat a \hat a^\dag$ operation in terms of equivalent input noise.
Using $\hat a \hat a^\dag$ and $(\hat a^{\dag})^2$ as basic building blocks, we compare combinatorial amplifications of coherent states using  $(\hat a \hat a^\dag)^2$, $\hat a^{\dag4}$, $\hat a \hat a^\dag\hat a^{\dag2}$, and  $\hat a^{\dag2}\hat a \hat a^\dag$, and show
that  $(\hat a \hat a^\dag)^2$, $\hat a^{\dag2}\hat a \hat a^\dag$, and  $\hat a^{\dag4}$  exhibit strongest noiseless properties for $|\alpha| \lesssim 0.51$, $0.51 \lesssim |\alpha| \lesssim 1.05 $, and $|\alpha|\gtrsim 1.05 $, respectively. We further show that the $(\hat a^{\dag})^2$ operation can be useful for amplifying superpositions of the coherent states.
In contrast to previous studies, our work provides efficient schemes to implement a noiseless amplifier for light fields with medium and large amplitudes.
\end{abstract}



\section{Introduction}

Quantum noise is a fundamental property which forbids quantum cloning \cite{Wootters:1982ex}, superluminal communication \cite{Herbert:1982br}, and violation of generalized uncertainty principle \cite{Arthurs:1988fv}.
Quantum theory prohibits deterministic linear amplification of  bosonic system to be noiseless \cite{Caves:1982hd}. The one way to circumvent this fundamental restriction is to adopt probabilistic amplification heralded by successful measurements \cite{Ralph:2008wi, Fiurasek:2009jg, Ferreyrol:2010, Xiang:2010hx, Zavatta:2011ea}.
Noiseless amplification can be used for many quantum information tasks such as entanglement distillation \cite{Xiang:2010hx}, continuous variable quantum key distribution \cite{Fossier:2009dz, Blandino:2012bg}, loss suppression \cite{Micuda:2012em}, quantum repeater \cite{Duan:2001dt}, phase estimation \cite{Usuga:2010fw}, quantum error correction \cite{Ralph:2011ct}, high-accuracy homodyne detection with low-efficiency detector \cite{Leonhardt:1994fm}, and quantum nonlocality tests \cite{Park:2012eq, Torlai:2013hw}.

In order to implement a nondeterministic noiseless amplifier, state-of-the-art  techniques of quantum optics are required. There are two types of noiseless linear amplification schemes for coherent states of small amplitudes. The first type proposed earlier \cite{Ralph:2008wi, Ferreyrol:2010, Xiang:2010hx} utilizes quantum scissors \cite{Pegg:1998} to implement a noiseless amplifier. The other employs the photon number operation $\hat n$ as its basic element \cite{Fiurasek:2009jg, Zavatta:2011ea}, where the photon subtraction \cite{Wenger:2004cw} and addition \cite{Zavatta:2004km} operations, represented by the photon annihilation and creation operators $\hat a$ and $\hat a^\dag$, are required for an  experimental implementation.
Recently, the first order approximation, $\hat a \hat a^{\dag}$, of the noiseless amplification was experimentally implemented \cite{Zavatta:2011ea}. 
This approach enables one to realize 
 a high-fidelity ($F > 0.9$) amplifier  with the fixed amplitude gain of $g=2$ for coherent states of very small amplitudes ($|\alpha| \lesssim 0.67$).
Therefore, it is important to develop an efficient amplification scheme for coherent states of larger amplitudes.

Another interesting issue is to apply a noiseless amplifier to superpositions of coherent states (SCSs). 
It is well known that the free-traveling SCSs of light with large amplitudes are useful for both fundamental studies of quantum mechanics such as Bell- and Legett-type inequality tests \cite{Mann:1995jy, Filip:2001gw, Wilson:2002fl, Wenger:2003eu, Jeong:2003es, Stobinska:2007co, Jeong:2008jy, Lee:2009kt, Gerry:2009gp, Jeong:2009dl, Paternostro:2010ca, Lee:2011kx, Lim:2012kj, Kirby:2013ev, Kirby:2014bd, Park:2015kl} 
 and for quantum information applications including precision measurements \cite{Gerry:2001ko, Gerry:2002hm, Ralph:2002ip, Munro:2002ew, Campos:2003ka, Joo:2011ge, Hirota:2011vl, Joo:2012fh, Zhang:2013jk}, quantum computation \cite{Cochrane:1999ck, Jeong:2002jj, Ralph:2003jk, Lund:2008hra, 
 Myers:2011ef, Kim:2015go}, quantum teleportation \cite{vanEnk:2001gb, Jeong:2001eu} 
 and quantum key distribution \cite{Simon:2014ih}. 
Squeezed single photon and squeezed vacuum states are often used to approximate SCSs \cite{Lund:2004tg,Jeong:2005kq, Ourjoumtsev:2006ima, Ourjoumtsev:2009jh}.
These implementations, however, are good approximations only for SCSs with small amplitudes and they exhibit low fidelities to SCSs with large amplitudes \cite{Lund:2004tg,Jeong:2005kq}. 
It is thus worth investigating whether the fidelities to large SCSs can be enhanced with the use of a nondeterministic noiseless amplifier.
We also note that a scheme for deterministic amplification of SCSs in circuit quantum electrodynamics was proposed \cite{Joo:2015wn} but this scheme cannot be applied to free-traveling SCSs.

In this paper, we show that the  two-photon addition ($(a^{\dag })^2$) works as a more effective noiseless amplifier, compared to the photon-addition-and-subtraction ($\hat a \hat a^{\dag}$), when amplifying  coherent states and SCSs of  relatively large amplitudes. Figures of merit examined here are the state fidelity, the amplitude gain, and the equivalent input noise (EIN) \cite{Grangier:1992fva}. The noiseless property of the amplification is assessed by the EIN of the amplifier \cite{Ferreyrol:2010, Zavatta:2011ea}, which is affected by both the state fidelity and the amplitude gain. 
Our analysis also shows that the amplified squeezed vacuum and single-photon states using the $(a^{\dag })^2$ operation exhibit higher fidelities to ideal SCSs than the states without the amplification and requires less squeezing for a large range of parameters.

The remainder of our paper is organized as follows. In Sec. \ref{Sec:Coherent state amplification}, we examine one- and two- cycle amplifications of coherent states. The fidelity, amplitude gain and noiseless property after the amplifications are investigated for comparisons. Sec. \ref{Sec:Amplification of superpositions of coherent states} is devoted to the amplification of SCSs. 
We conclude with final remarks in Sec. \ref{Sec:Conclusion}.

\section{Amplification of coherent states}
\label{Sec:Coherent state amplification}

\subsection{One-cycle amplification}
\label{Subsec:one-cycle amplification}

Applying the amplification operator $\hat A \in \{\hat{a} \hat{a}^\dag, \hat{a}^{\dag2}\}$ to a coherent state of amplitude ${\alpha_i}$, the amplified state is expressed as 
\begin{equation}
 N^{\hat A} ({\alpha_i}) \hat{A} \left| { {\alpha_i}} \right\rangle,
\label{eq:SingleNormalization}
\end{equation}
where the normalization factors are 
\begin{align}
	N^{\hat a\hat{a}^\dag} ({\alpha_i})= ({{\alpha_i}^4} + 3{{\alpha_i} ^2} + 1)^{-\frac{1}{2}},\\
	N^{\hat{a}^{\dag2}} ({\alpha_i})=({{\alpha_i}^4} + 4{{\alpha_i} ^2} + 2)^{-\frac{1}{2}},
\label{eq:SingleNCA}
\end{align} 
and $\alpha_i$ is  assumed to be  real without loss of generality.

The fidelity between the $\hat A$-amplified coherent state of initial amplitude ${\alpha_i}$ and the coherent state of  real amplitude $\alpha_f$ is  
\begin{equation}
	{F^{\hat A}}=   {\{N^{\hat A} ({\alpha_i})\}^2}{\left| \left\langle {  {\alpha_f} } \right|
  \hat{A} \left| { {\alpha_i}} \right\rangle  \right|^2}.
\label{eq:SingleFC}
\end{equation}
We have obtained analytic expressions of ${F^{\hat A}}$ as
\begin{align}
&	{F^{\hat a\hat{a}^\dag}}
	={ \left\{ N^{\hat a\hat{a}^\dag} ({\alpha_i})\right\}^2}{e^{-(\alpha_f-{\alpha_i})^2}   (\alpha_f  {\alpha_i} +1)^2},\\
&	 {F^{\hat{a}^{\dag2}}}={\left\{ N^{\hat{a}^{\dag2}}({\alpha_i})\right\}^2}{ \alpha_f^4 e^{-( {\alpha_f} -\alpha _i)^2}}.
	\label{eq:SingleFCA}
\end{align}

\begin{figure}[t]
	\begin{centering}
		\includegraphics[width=10cm]{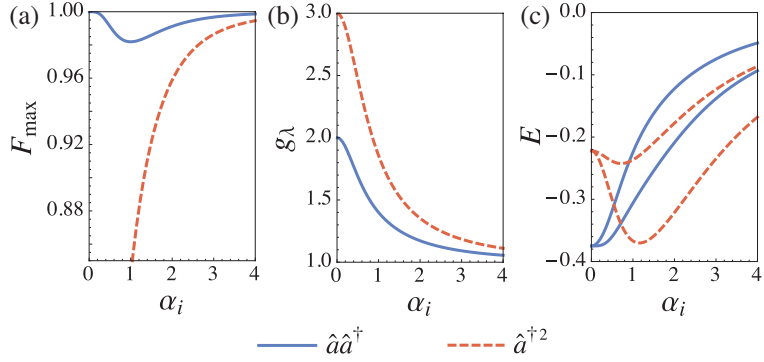}
			\caption{(a) Maximum fidelities, (b) amplitude gains, and  (c) EINs when the amplification methods $\hat a \hat a^\dag$ (solid curve) and $\hat a^{\dag2}$ (dashed curve)  are applied to the coherent state of initial  amplitude $\alpha_i$. (a) The fidelities of the $\hat a \hat a^\dag$-amplified coherent states are higher than $F_{\rm max}>0.98$, which are close to 1 for small and large $\alpha_i$. The fidelity of the $\hat a^{\dag2}$-amplified coherent state approaches 1 for large $\alpha_i$. (b) The higher amplitude gain is obtained when the amplification is performed by  $\hat a^{\dag2}$ rather than  $\hat a \hat a^\dag$. The gains from $\hat a \hat a ^\dag$ and $\hat a ^{\dag2}$ approach 2 and 3 as $\alpha_i\to0$, and are dropped to 1 as $\alpha_i\to \infty$. (c) The upper solid and dashed curves represent average EINs, i.e. EINs averaged over all values of $\lambda$, while the lower solid and dashed curves correspond to EINs with $\lambda=0$, which gives the lowest EINs. The $\hat a^{\dag2}$-amplification exhibits lower average EINs than the $\hat a \hat a^\dag$-amplification with large  amplitude  $\alpha_i\gtrsim0.91$, while the opposite is true for $\alpha_i\lesssim0.91$. As $\alpha_i \to 0$, the average EINs approach $-{3}/{8}$ for $\hat a \hat a^\dag$, and $-{2}/{9}$ for $\hat a^{\dag2}$. The average EINs approach zero as $\alpha_i$ increases for both the cases.
			}
		\label{fig:FigCFirst}
	\end{centering}
\end{figure}

We  take the maximum fidelity of  the $\hat A$-amplified coherent state of initial amplitude $\alpha_i$ to the final coherent state as
\begin{equation}
	F^{\hat A}_{\rm max}(\alpha_i)= 
	\mathop {\max } \limits_{{\alpha_f}}	{F^{\hat A}}
	\label{eq:maxSingleFC}
\end{equation}
where the maximum is taken over the amplitude $\alpha_f$.
The results of numerical maximization by steepest descent \cite{numericalrecipes} are presented in Fig.~\ref{fig:FigCFirst}(a).
The $(\hat a \hat a^\dag)$-amplification always exhibits higher maximum fidelity than the $(\hat a^\dag)^2$-amplification.
On the other hand, it is the opposite for the amplitude gain as explained in what follows. 
The amplitude gain from the amplification $\hat A$ can be defined as the ratio of the expectation values of the quadrature operator with phase $\lambda$ ($\hat x_\lambda$) \cite{Zavatta:2011ea}:
\begin{equation}
g^{\hat A}_\lambda =\frac{|N^{\hat A} ({\alpha_i})|^2{\left| \left\langle {\alpha_i}  \right| \hat A^\dag  \hat x_\lambda \hat A \left| {\alpha_i}\right\rangle \right| }}{{\left| \left\langle {\alpha_i}  \right|  \hat x_\lambda  \left| {\alpha_i}\right\rangle \right| } }, 
\label{eq:gainC}
\end{equation}
of which explicit expressions are obtained as
\begin{align}
&{g^{\hat a\hat{a}^\dag}_\lambda}={ \left\{ N^{\hat a\hat{a}^\dag} ({\alpha_i})\right\}^2} (\alpha_i ^4+4 \alpha_i^2 + 2),\\
&{g^{\hat{a}^{\dag2}}_\lambda}={\left\{ N^{\hat{a}^{\dag2}}({\alpha_i})\right\}^2}{ \left(\alpha_i ^4+6 \alpha_i ^2+6\right)}.
\label{eq:SinglegainCA}
\end{align}
The amplitude gains are independent of $\lambda$ for the amplifications of coherent states considered here. The gain monotonically decreases to unity with respect to $\alpha_i$ (Fig.~\ref{fig:FigCFirst}(b)), since $\hat{a} \hat{a}^\dag$ or ${\hat{a}^{\dag2}}$ merely alters the ratios of the superposition of Fock states for large-amplitude coherent states. The maximum fidelity in Fig.~\ref{fig:FigCFirst}(a) reaches unity for large $\alpha_i$ with the same reason.

We now employ the equivalent input noise (EIN) \cite{Grangier:1992fva} for comparison between the two amplification schemes.  The EIN came from a classical electronics terminology used to quantify the performance of  an amplifier considering the amplification gain and the generated noise at the same time. It measures the amount of noise that must be added to the input noise level to mimic the observed output noise for the given gain using a classical amplifier \cite{Ferreyrol:2010}. When a quantum amplifier is used, the EIN can be negative
\cite{Ferreyrol:2010,Zavatta:2011ea}. 
The EIN of an amplifier is defined as 
\begin{equation}
 E_\lambda^{\hat A}=\frac{\left\langle \Delta x_\lambda \right\rangle^2_{\rm out}}{(g^{\hat A}_\lambda)^2}- \left\langle \Delta x_\lambda \right\rangle^2_{\rm in},
\label{eq:EIN}
\end{equation}
where
$\left\langle \Delta x_\lambda \right\rangle_{\rm in}^2$ and $\left\langle \Delta x_\lambda \right\rangle_{\rm out}^2$ are the expectation values of the quadrature variance operator with phase $\lambda$ for input and output states.   
 The first term of the right hand side of Eq. (\ref{eq:EIN}) represents the level of the input noise to classically mimic the output noise, and the second term represents the actual level of input noise. Thus the difference between the two terms corresponds to the level of noise added into the input signal to mimic the output signal in quadrature $x_\lambda$.
We find its explicit forms as 
\begin{equation}
\begin{split}
E_\lambda^{\hat a \hat a^\dag} (\alpha_i)&= \frac{ \left\{ 2 {\alpha_i} ^6+11 {\alpha_i} ^4+11 {\alpha_i} ^2+2 \left({\alpha_i} ^4+5 {\alpha_i} ^2+3\right) {\alpha_i} ^2 \cos (2 \lambda )+1\right\}}{2 { \{ N^{\hat a \hat{a}^{\dag}} ({\alpha_i})\}^2} \left({\alpha_i} ^4+4 {\alpha_i} ^2+2\right)^2}- 2 {\alpha_i} ^2 \cos^2  \lambda-\frac{1}{2}\\
E_\lambda^{ \hat a^{\dag2}} (\alpha_i)&=
\frac{\{ 2 N^{\hat{a}^{\dag2}} ({\alpha_i})\}^2 \left(\alpha ^2+2\right) \left\{\alpha ^4+\left(\alpha ^2+6\right) \alpha ^2 \cos (2 \lambda )+6 \alpha ^2+2\right\}+1}{{ \{ 2 N^{\hat{a}^{\dag2}} ({\alpha_i})\}^4}\left(\alpha ^4+6 \alpha ^2+6\right)^2}\\
&-2 \alpha ^2 \cos ^2 \lambda -\frac{1}{2}
\end{split}
\label{eq:EINC}
\end{equation}
The EINs with $\lambda=0$ ($ E^{\hat a \hat a ^\dag}_0 (\alpha_i)$ and $E^{a^{\dag2}}_0 (\alpha_i)$) and the $\lambda$-averaged EINs ($ E^{\hat a \hat a ^\dag} (\alpha_i)$ and $E^{a^{\dag2}} (\alpha_i)$) are  plotted in Fig. \ref{fig:FigCFirst}(c). The parameter $\lambda=0$ is chosen because it gives  lower value of EIN than any other $\lambda$.
All average EINs are negative, which indicates the characteristic of noiseless amplification; negative EIN cannot be obtained by the classical amplification. As the $\hat a \hat a^\dag$-amplification has much higher fidelity than $\hat a^{\dag2}$ for small $\alpha_i$, $\hat a \hat a^\dag$ has lower EIN for small $\alpha_i$, while  $\hat a^{\dag2}$ has lower EIN for large $\alpha_i$ due to higher amplitude gains.

\begin{figure}[t]
	\begin{centering}
		\includegraphics[width=10cm]{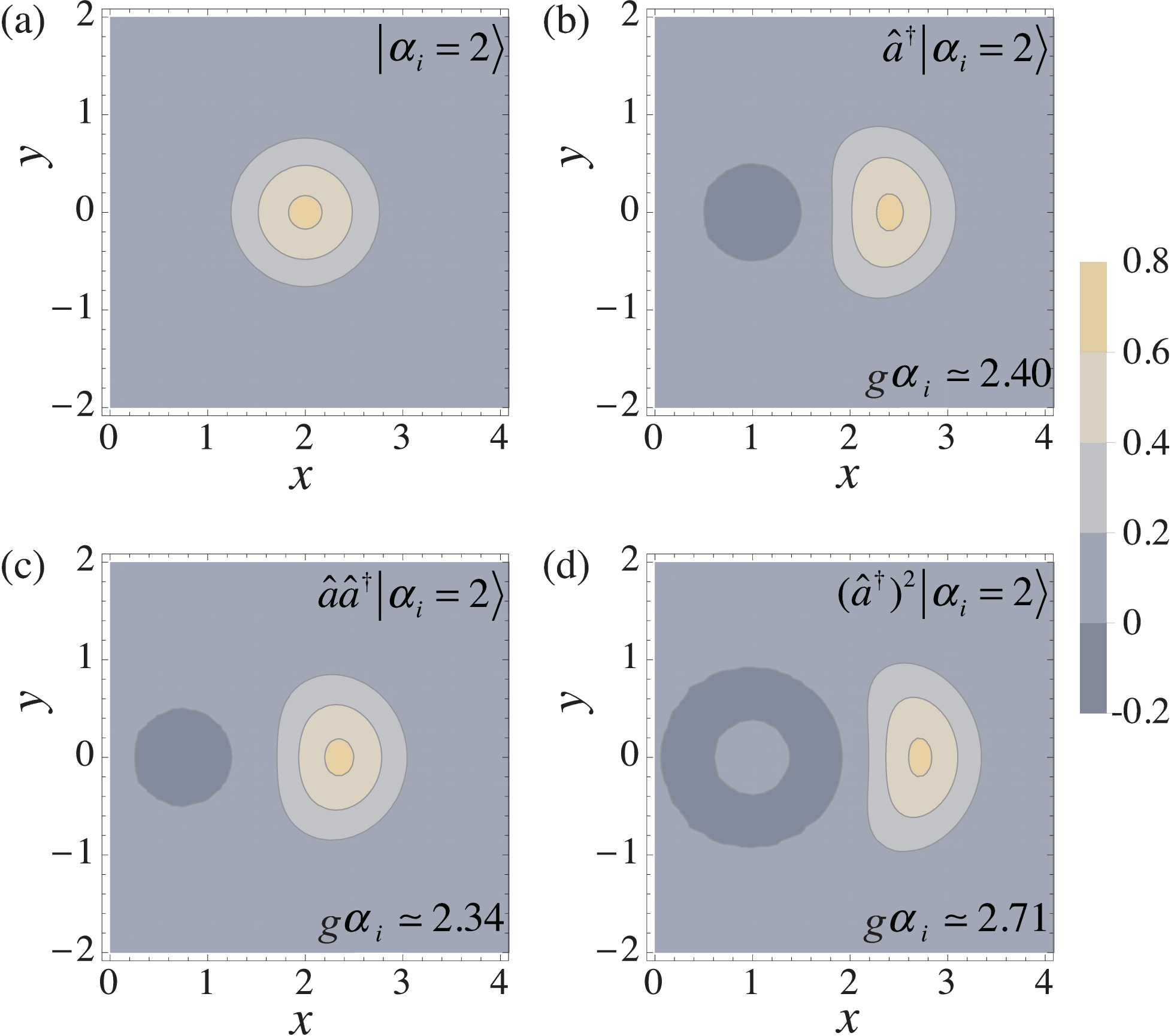}
			\caption{Wigner functions of (a) a coherent state of amplitude $\alpha_i=2$ and the states after photonic operations, (b) $\hat a^\dag $, (c) $\hat a \hat a^\dag$, and (d) $(\hat a^\dag)^2$.  The amplitudes of the amplified states, $g \alpha_i$, are noted for comparison.
			}
		\label{fig:wigner}
	\end{centering}
\end{figure}

The $\hat{a}\hat{a}^\dag$-amplification of the coherent state shows higher fidelities to the coherent states than the $\hat a^{\dag2}$-amplification. However, when the initial amplitude is large enough to approach a sufficiently high fidelity, the $\hat a^{\dag2}$-amplification is advantageous in terms of amplitude gain and EIN.

In order to better understand how the amplification processes works, we plot changes of the Wigner functions \cite{Wigner1932} of coherent states after the photonics operations in Fig. \ref{fig:wigner}. 
When the operation $\hat a^\dag$ is applied to a coherent state of $\alpha_i=2$, the peak of the Wigner function  moves from the origin while its shape becomes less circular than the original coherent state as shown in Fig. \ref{fig:wigner}(b). If the operation $\hat a$ is successively applied to $\hat a^\dag | \alpha_i \rangle$, the shape of the Wigner function (Fig. \ref{fig:wigner}(c)) becomes more circular than $\hat a^\dag | \alpha_i \rangle$, but the distance from the origin somewhat decreases. In other words, the final state better approximates a coherent state at the price of less amplification. 
However, if the operation $\hat a^\dag$ is applied, instead of  the second operation $\hat a$, to the state $\hat a^\dag | \alpha_i \rangle$, the Wigner function becomes a little more distorted than $\hat a^\dag | \alpha_i \rangle$, but its amplitude becomes larger as shown in Fig. \ref{fig:wigner}(d). We provide explicit forms of the Wigner functions in Appendix A.

\subsection{Two-cycle amplification}
\label{Subsec:Two-cycle amplification}

\begin{figure}[t]
	\begin{centering}
		\includegraphics[width=10cm]{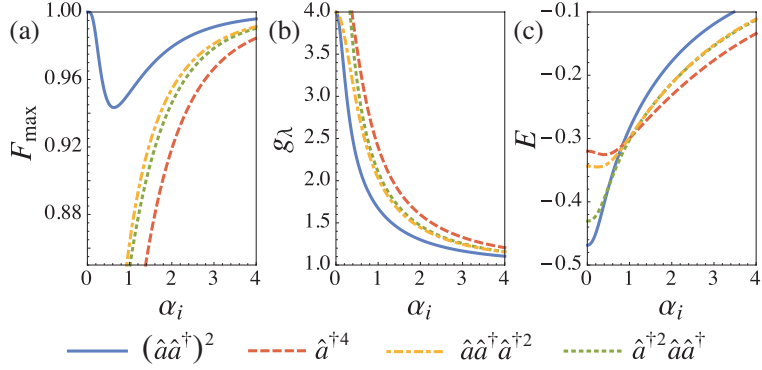}
		\caption{(a) Maximum fidelities, (b) amplitude gains, and  (c) average EINs after two-cycle amplifications $(\hat{a} \hat{a}^\dag)^2$ (solid curve), $\hat{a}^{\dag4}$ (dashed curve),  $\hat{a} \hat{a}^\dag\hat{a}^{\dag2}$ (dot-dashed curve), and $\hat{a}^{\dag2}\hat{a} \hat{a}^\dag$ (dotted curve). (a) The maximum fidelity of the $(\hat a \hat a^\dag)^2$-amplified coherent state approaches 1 for small and large $\alpha_i$'s, which is the highest among the two-cycle amplifications. All the maximum fidelities become perfect as the initial amplitude $\alpha_i$ increases. (b) The higher gain is obtained by the two-cycle amplifications compared to the one-cycle amplifications. The gains from $(\hat{a} \hat{a}^\dag)^2$, $\hat{a}^{\dag4}$, $\hat{a}^{\dag2}\hat{a} \hat{a}^\dag$, and $\hat{a} \hat{a}^\dag\hat{a}^{\dag2}$ become 4, 5, 4, and 6 as $\alpha_i\rightarrow0$, respectively, and drop to 1 as $\alpha_i$ increases. (c) In the regions of $\alpha_i \lesssim 0.51 $, $0.51 \lesssim \alpha_i \lesssim 1.05 $, and $ \alpha_i\gtrsim1.05$,  $(\hat a \hat a^\dag)^2$, $\hat a^{\dag4}$, and  $\hat a^{\dag2}\hat a \hat a^\dag$ show the lowest EIN, respectively, which are all lower than EINs obtained by the one-cycle amplifications. }
	\label{fig:FigCSecond}
	\end{centering}
\end{figure}

We consider four possible combinatorial two-cycle amplifications, 
$\hat A \in \{(\hat{a} \hat{a}^\dag)^2, \hat{a}^{\dag4}, $ $\hat{a}^{\dag2}\hat{a} \hat{a}^\dag, \hat{a} \hat{a}^\dag\hat{a}^{\dag2}\}$, where each of the four processes are a combination from the two basic amplification units, $\hat a \hat a^\dag$ and $(\hat a^{\dag})^2$. We have obtained the  fidelities, gains and EINs
as explained in the previous section using Eqs. (4), (8) and (11), and the results are presented in Appendix B. The maximum fidelity $F^{\hat A}_{\rm max}(\alpha_i)$ in terms of the initial amplitude $\alpha_i$ is numerically obtained \cite{numericalrecipes} and plotted in Fig.~\ref{fig:FigCSecond}(a). Among the two-cycle amplifications, the $(\hat{a} \hat{a}^\dag)^2$-amplification exhibits the highest maximum fidelity to the coherent state, although the fidelity is slightly lower than one-cycle amplification, $\hat{a} \hat{a}^\dag$. 
The order of the fidelity performance is 
\begin{equation}
F^{(\hat a \hat a ^\dag)^2}_{\rm max} (\alpha_i) > F^{\hat a \hat a ^\dag \hat a^{\dag2}}_{\rm max} (\alpha_i) > F^{\hat a^{\dag2}\hat a \hat a ^\dag}_{\rm max} (\alpha_i)  > F^{\hat a^{\dag4}}_{\rm max} (\alpha_i).
\end{equation}

All the gains from the two-cycle amplifications (Eq. (\ref{eq:twocyclegainCA}) of Appendix B) are higher than those from one-cycle amplification, which also monotonically decreases to unity with respect to $\alpha_i$ (Fig.~\ref{fig:FigCSecond}(b)). The gain from $(\hat{a} \hat{a}^\dag)^2$ is the lowest among the two-cycle amplifications, although the fidelity is the highest. For sufficiently large amplitude ($  \alpha_i\gtrsim0.27$), the following relation holds for amplitude gains:
\begin{equation}
g^{\hat a^{\dag4}}_{\lambda} (\alpha_i) > g^{\hat a^{\dag2}\hat a \hat a ^\dag}_{\lambda} (\alpha_i) >  g^{\hat a \hat a ^\dag \hat a^{\dag2}}_{\lambda} (\alpha_i) > g^{(\hat a \hat a ^\dag)^2}_{\lambda} (\alpha_i).
\end{equation}
Integrating EINs (Eqs. (\ref{eq:twocycleEINC}) and (\ref{eq:twocycleEINC2})) to obtain $\lambda$-averaged EINs,  $E^{(\hat a \hat a ^\dag)^2} (\alpha_i)$, $E^{\hat a^{\dag2}\hat a \hat a ^\dag} (\alpha_i)$, $E^{\hat a \hat a ^\dag \hat a^{\dag2}} (\alpha_i)$, and $E^{\hat a^{\dag4}} (\alpha_i)$  are plotted in Fig. \ref{fig:FigCSecond}(c), which are all negative indicating the characteristic of noiseless amplification. The following two-cycle amplifications achieve the lowest EINs, including one-cycle amplifications, in the corresponding regions: $(\hat a \hat a^\dag)^2$ in $\alpha_i \lesssim 0.51 $, $\hat a^{\dag4}$ in $0.51 \lesssim \alpha_i \lesssim 1.05 $, and  $\hat a^{\dag2}\hat a \hat a^\dag$ in $\alpha_i\gtrsim1.05$.

\subsection{Success probabilities of the amplification processes}

The photon subtraction process uses a beam splitter and a photodetector while the photon addition process relies on a parametric down converter and a photodetector
\cite{Zavatta:2004km}. 
The success probability of the photon subtraction thus depends on the ratio of the beam splitter, while the success probability of the photon addition is determined by the parametric gain.
It should be noted that the reflectivity of the beam splitter should be very small in order to well approximate the photon annihilation operator $\hat{a}$. 
Assuming an ideal single-photon detector, the success probability of photon addition using a parametric down converter with parametric gain $\lambda_g\ll1$ is \cite{Bellini:2010}
\begin{align}
p_{add} &\approx |\lambda_g|^2 (\overline n +1),
\end{align}
where $\overline n$ is the average photon number of the initial state. 
The success probability of photon subtraction using a beam splitter of reflectivity $R \ll1$ is approximated as
\begin{align}
p_{sub} &\approx R  \overline n.
\end{align}
For instance, with realistic experimental parameters of $\lambda_g\sim  0.1$ \cite{nf01,nf02} and $R\sim 0.05$ the realization of $\hat a \hat a^\dag$ and $(\hat a^\dag)^2$ to noiselessy amplify a coherent state of amplitude $\alpha_i = 2$ have estimated probability of $\sim 10^{-2}$ and $\sim 10^{-3}$, respectively. Such probabilities are very well compatible with the realization of the proposed schemes with pulsed pump lasers where the final heralding rate is enhanced by the high pulse repetition rate of the laser $\sim 10^8 $ Hz \cite{Zavatta:2011ea,Zavatta:2004km,Parigi:2007}. However, the implementation of the $(\hat a^\dag)^4$, $\hat a \hat a^\dag \hat a^{\dag 2}$, and $\hat a^{\dag 2} \hat a \hat a^\dag $ with the same experimental parameters of above will results with a probability of $\sim 10^{-5}$ which is more demanding with the present technology.

\section{Amplification of superpositions of coherent states}
\label{Sec:Amplification of superpositions of coherent states}

\subsection{Ideal even and odd superpositions of coherent states}

\begin{figure}[t]
	\begin{centering}
		\includegraphics[width=13cm]{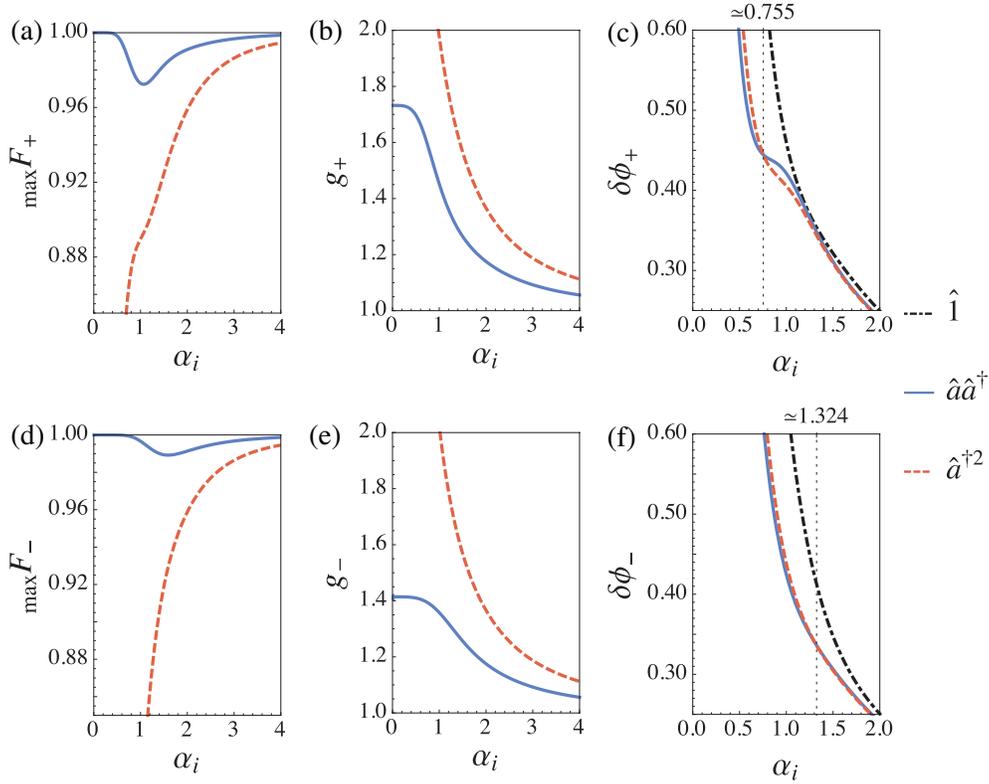}
			\caption{(a) Maximum fidelities, (b) amplitude gains, and  (c) optimal phase uncertainties when the amplification methods $\hat a \hat a^\dag$ (solid curve) and $\hat a^{\dag2}$ (dashed curve)  
are applied to the even SCS of initial  amplitude $\alpha_i$. The same functions for the case of the odd SCS are plotted in panels (d), (e) and (f). The dot-dashed curves in panels (c) and (f) represent  optimal phase uncertainties of the even and odd SCSs, respectively.
The amplitude gains using $\hat a \hat a ^\dag$ approach $\sqrt 3$ and $\sqrt 2$ for even and odd SCSs, respectively, as the initial amplitude $\alpha_i\to0$. The amplitude gain using either of the amplification methods approaches 1 as $\alpha_i\to \infty$.
			}
		\label{fig:Cat}
	\end{centering}
\end{figure}

Even and odd SCSs are defined as 
\begin{align}
	\left| {{\pm }_\alpha } \right\rangle  = \frac{1}{\sqrt{2(1\pm e^{-2|\alpha|^2} )}}(\left| \alpha  \right\rangle  \pm \left| { - \alpha } \right\rangle ),
	\label{eq:cpm}
\end{align}
where $\left| {+_\alpha } \right\rangle$ ($\left| {-_\alpha } \right\rangle$) certainly contains an even (odd) number of photons as implied by its name.
The amplified SCSs with the initial amplitude $\alpha_i$ are
\begin{align}
	| {{\pm_{\alpha_i}^{\hat A} }} \rangle ={N_{\pm}^{\hat A} (\alpha_i)} \hat{A} \left| { \pm_{\alpha_i}} \right\rangle,
\label{eq:AmplifiedSCS}
\end{align}
where
\begin{equation}
\begin{split}
	N_\pm^{\hat a\hat{a}^\dag} (\alpha_i)&= [2\{   \pm {e^{ - 2{\alpha_i ^2}}}\left( {{\alpha_i ^4} - 3{\alpha_i ^2} + 1} \right) +({\alpha_i ^4} + 3{\alpha_i ^2} + 1)\}]^{-\frac{1}{2}},
\\
	N_\pm^{\hat{a}^{\dag2}} (\alpha_i)&=[2\{  \pm {e^{ - 2{\alpha_i ^2}}}\left( {{\alpha_i ^4} -4{\alpha_i ^2} + 2} \right) +({\alpha_i ^4} + 4{\alpha_i ^2} + 2)\}]^{-\frac{1}{2}}
\label{eq:SCSN}
\end{split}
\end{equation} 
and $\alpha_i$ is assumed to be real without loss of generality.
  The maximum fidelities of the $\hat A$-amplified even and odd SCSs with initial amplitude $\alpha_i$ are
\begin{equation}
	_{\rm max}F_{\pm}^{\hat A}(\alpha_i) = \mathop {\max } \limits_{\alpha_f}  {\left| {\left\langle {  \pm_{\alpha_f} }
\Big|
 \pm_{\alpha_i}^{\hat A} \right\rangle } \right|^2},
	\label{eq:FmaxSCS}
\end{equation}
where 
\begin{equation}
\begin{split}
	{F_\pm^{\hat a\hat{a}^\dag}}
	&=\frac{2 \left\{ N_\pm^{\hat a\hat{a}^\dag} (\alpha_i)\right\}^2 e^{-(\alpha_f+\alpha_i)^2}  \left\{e^{2 \alpha_f  \alpha_i } (\alpha_f  \alpha_i +1)\mp(\alpha_f  \alpha_i -1)\right\}^2}{\pm e^{-2 \alpha_f^2} +1},\\
{F_\pm^{\hat{a}^{\dag2}}}&=\frac{2 \left\{ N^{\hat{a}^{\dag2}}_\pm(\alpha_i)\right\}^2 \alpha_f^4 \left\{\pm e^{-\frac{1}{2}( \alpha_i +\alpha_f )^2}+e^{-\frac{1}{2}( \alpha_i -\alpha_f)^2}\right\}^2}{\pm e^{-2 \alpha_f^2} +1}.
	\label{eq:SCSF}
\end{split}
\end{equation}
We observe from the numerical results \cite{numericalrecipes} plotted in Figs.~\ref{fig:Cat}(a) and \ref{fig:Cat}(d)  that
the $(\hat a \hat a^\dag)$-amplification results in higher maximum fidelities than the $(\hat a^\dag)^2$-amplification. The maximum fidelities of the $\hat a \hat a^\dag$-amplified even and odd SCSs are higher than $F_{\rm max}>0.97$ (Fig.~4(a)) and $F_{\rm max}>0.98$ (Fig.~4(d)), respectively. Clearly, the maximum fidelities using $\hat a \hat a^\dag$ approach 1 for small and large values of $\alpha_i$ while those using $\hat a^{\dag2}$ approach 1 only for large values of $\alpha_i$.

Unlike coherent states, which has Gaussian probability distributions in the measurement of $\hat x_{\lambda}$, the  definitions of the amplitude gain in Eq. (\ref{eq:gainC}) and the EIN in Eq. (\ref{eq:EIN}) cannot be applied to the even and odd SCSs. We first define the amplitude gain as the ratio between the input amplitude of SCS and the output amplitude of SCS that maximizes the fidelity as
\begin{equation}
g^{\hat A}_{\pm} (\alpha_i) = \frac{|\alpha_f|}{|\alpha_i|},
\end{equation}
where $\alpha_f$ maximizes $F_{\pm}^{\hat A}$. 
 When this definition is applied to the case of coherent states, we do not obtain exactly the same results with those obtained using Eq.~(8), although the differences become negligible when the fidelities are high.
As shown in Figs.~\ref{fig:Cat}(b) and  \ref{fig:Cat}(e), it is numerically \cite{numericalrecipes} verified  that  the gain from $\hat a^{\dag2}$ is always higher than that from $\hat a \hat a^\dag$. 

 It is a nontrivial task to develop an equivalent notion and definition of the EIN for SCSs because the definition of noise for SCSs is not clear in this context. We pay attention to the fact that the larger SCSs are more useful for phase estimation \cite{Joo:2011ge}. Largely amplified SCSs with high fidelities should become more useful for phase estimation. In fact, the optimal phase estimation is closely related to both of the amplitude gain and the noiseless property of amplifiers \cite{Joo:2011ge}, and these two quantities are what the EIN quantifies.  We thus compare the optimal phase estimations obtained from Cram\'{e}r-Rao bound \cite{Braunstein:1994ug, Braunstein:1996ke} for even and odd SCSs,
\begin{equation}
\delta \phi_{\pm} = \frac{1}{\sqrt {{\cal F}_{\pm}}},
\label{eq:CRbound}
\end{equation} 
where ${\cal F}_{\pm}$ is quantum Fisher information
\begin{equation}
{\cal F}_{\pm} = 4 \langle (\Delta \hat n )^2 \rangle
\end{equation}
for pure states.

The analytic results for $\hat A \in \{\hat 1,~ \hat{a} \hat{a}^\dag,~ \hat{a}^{\dag2}\}$ are presented in Appendix C, where $\hat{1}$ denotes the identity operator.
The enhancement in phase uncertainty obtained by the two amplification schemes, 
 $\hat a^{\dag2}$ and $\hat a \hat a^\dag$, are shown in Figs.~\ref{fig:Cat}(c) and \ref{fig:Cat}(f) 
   as signatures of noiseless amplifications.
               As expected from the case of coherent states,  the $\hat a^{\dag2}$-amplification is more efficient for phase estimation with SCSs of  large amplitudes  than the $\hat a \hat a^\dag$-amplification.
The $\hat a^{\dag2}$ amplification exhibits lower phase uncertainties (i.e., better for phase estimation) than  $\hat a \hat a^\dag$ when applied to  even and odd SCSs with  large amplitude ($\alpha_i\gtrsim0.755$ and $\alpha_i\gtrsim1.324$, respectively), while the opposite is true for smaller $\alpha_i$.   
 It is also clear in Figs.~4(c) and 4(f) that the phase uncertainties decrease after $\hat a^{\dag2}$ and $\hat a \hat a^\dag$ are applied to the even and odd SCSs of amplitude $\alpha_i$, respectively.
 
\subsection{Approximations with squeezed vacuum and squeezed single-photon states}

\begin{figure}[t]
\begin{centering}
\includegraphics[width=13cm]{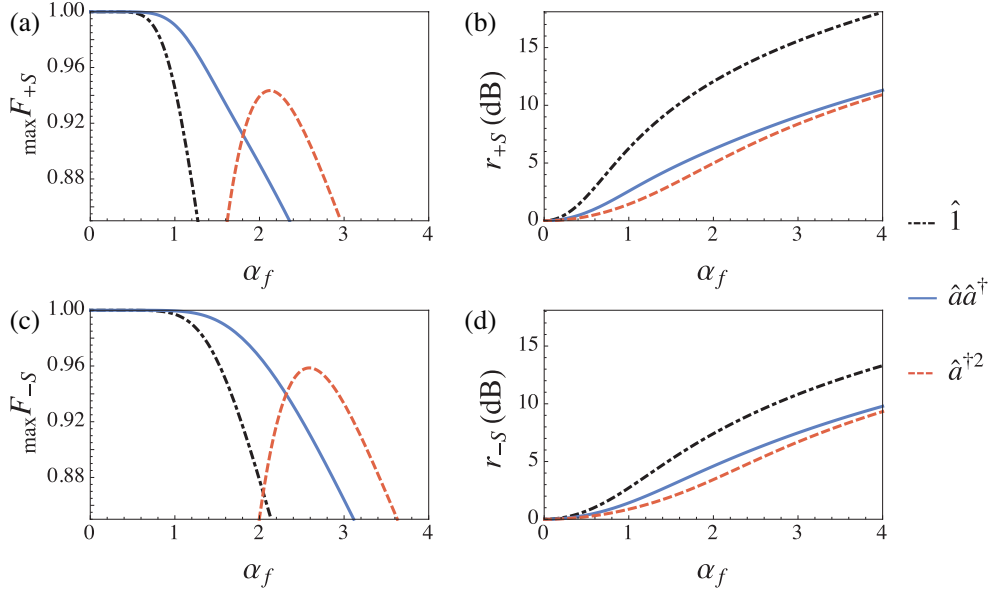}
\caption{(a) Maximum fidelities and  (b)  required levels of squeezing (dB) when the amplification methods $\hat a \hat a^\dag$ (solid curve) and $\hat a^{\dag2}$ (dashed curve)  
are applied to the squeezed vacuum state to approximate the even SCS of amplitude $\alpha_f$. The same functions  for the cases of the squeezed single-photon state 
for approximating the odd SCS are plotted in panels (c) and (d). The dot-dashed curves represent the cases with the squeezed vacuum and single-photon states, respectively, without the amplification methods.
In (a) and (c), higher fidelities are obtained for approximating the even and odd SCSs of large amplitudes 
 using the amplification methods $\hat a \hat a^\dag$ and $\hat a^{\dag 2}$ ($\alpha_f\gtrsim1.47$ and $\alpha_f\gtrsim2.04$, respectively), compared to the cases without the amplification methods. 
The amplification method  $\hat a^{\dag2}$ achieves the fidelities to even and odd SCSs up to $_{\rm max}F_{+S}^{ \hat a ^{\dag2}}\simeq0.943$ and $_{\rm max}F_{-S}^{ \hat a^{\dag2}}\simeq0.959$ around $\alpha_f\simeq2.12$ and $\alpha_f\simeq2.59$, respectively.
}
\label{fig:SqueezedFR}
\end{centering}
\end{figure}

It is known that a squeezed vacuum state and a squeezed single-photon state well approximate the even and odd SCSs with small amplitudes, respectively \cite{Lund:2004tg, Jeong:2005kq}, and that  multiple applications of the photon addition on the squeezed vacuum produces a squeezed SCS of a very high fidelity \cite{Marek08}. 
The squeezed vacuum and squeezed single-photon states can be expressed in the number state basis as
\begin{align}
		\hat S(r)\left| {0} \right\rangle &= \sum\limits_{n = 0}^\infty  {\frac{{{{(-\tanh r)}^n}}}{{{{(\cosh r)}^{ 1/2}}}}\frac{{\sqrt {(2n)!} }}{{{2^n}n!}}} \left| {2n} \right\rangle,\\
		\hat S(r)\left| 1 \right\rangle &= \sum\limits_{n = 0}^\infty  {\frac{{{{(-\tanh r)}^n}}}{{{{(\cosh r)}^{ 3/2}}}}\frac{{\sqrt {(2n + 1)!} }}{{{2^n}n!}}} \left| {2n + 1} \right\rangle,
\end{align}
where $\hat{S} (r)= \exp[-r (\hat{a}^2-\hat{a}^{\dag 2})/2]$ is the squeezing operator with squeezing parameter $r$. When an even (odd) SCS of amplitude $\alpha_f$ is desired,  the maximum fidelity which the squeezed vacuum (single-photon) state can achieve is
\begin{align}
_{\rm max}F_{ +S}^{\hat 1} (\alpha_f) &=  \mathop {\max }\limits_{r} \big| \langle { +_{\alpha_f} } |  \hat S(r)\big| {0} \rangle   \big|^2,\\
_{\rm max}F_{-S}^{\hat 1} (\alpha_f) &=  \mathop {\max }\limits_{r} \big| \langle { -_{\alpha_f} } |  \hat S(r)\big| {1} \rangle   \big|^2.
\label{eq:spm}
\end{align}
The dot-dashed  curve in Fig.~\ref{fig:SqueezedFR}(a) (Fig.~\ref{fig:SqueezedFR}(c)) shows that the maximum fidelity of the squeezed vacuum (single-photon) state to the even (odd) SCS of amplitude $\alpha_f$ approaches unity, as $\alpha_f \to 0$.

Now, we apply the amplifier 
$\hat A \in \{\hat{a} \hat{a}^\dag, \hat{a}^{\dag2}\}$ to the squeezed vacuum and single-photon states, $\hat S(r)\left| {0} \right\rangle$ and $\hat S(r)\left| {1} \right\rangle$ in order to obtain approximate SCSs.
The amplified squeezed vacuum and single-photon states are
\begin{align}
| + S_r^{\hat A}  \rangle &=M_+^{\hat A} (r) \hat A  \hat S(r)\left| {0} \right\rangle,\\
| - S_r^{\hat A}  \rangle &=M_-^{\hat A} (r) \hat A  \hat S(r)\left| {1} \right\rangle, 
\label{eq:AmplfiedSqueezed}
\end{align}
where 
\begin{equation}
\begin{split}
M_{\pm}^{\hat{a} \hat{a}^\dag} (r)&=
{({\rm sech} [r]})^2{\Big [ \frac{\{(1\mp 1) \tanh ^4r+(12\mp 8 ) \tanh ^2r+{5\mp3 }\}}{2} \
\Big]^{-\frac{1}{2}}},
\\
M_{\pm}^{\hat{a}^{\dag2}} (r)&={({\rm sech} [r]})^2{ [{(5\mp4) \tanh ^2r+4\mp2}]^{-\frac{1}{2}}}.
\label{eq:SqN}
\end{split}
\end{equation}
The fidelities between the $\hat A$-amplified squeezed vacuum (single-photon) state with initial squeezing $r$ and the ideal even (odd) SCS of amplitude $\alpha_f$ are
\begin{align}
  F_{\pm S}^{\hat a \hat a ^\dag}&={\big| {\langle { \pm_{\alpha_f} }
|
  \pm S_r^{\hat a \hat a ^\dag} \rangle } \big|^2}=\frac{2 \{M_\pm^{\hat a \hat a ^\dag} (r)\}^2 \alpha_f ^{1\mp1} e^{-\alpha_f^2 (\tanh r-1)}   \big(\alpha_f^2 \tanh r-\frac{3\mp1}{2}\big)^2}{ \big(\pm 1+ e^{2 \alpha_f ^2}\big) \text{cosh}^{2\mp1}r},
\\
F_{\pm S}^{ \hat a ^{\dag2}}&={\big| {\langle { \pm_{\alpha_f} }|
  \pm S_r^{\hat a ^{\dag2}} \rangle } \big|^2}=\frac{2 \{ M_{\pm}^{\hat  a ^{\dag2}} (r)\}^2 \alpha_f ^{5\mp1} e^{-\alpha_f ^2 (\tanh r+1)}  }{ \big(1\pm e^{-2 \alpha_f ^2} \big)\text{cosh}^{2\mp1}r},
\end{align}
where subscript `$+$' (`$-$') denotes the even (odd) SCS.
The maximum fidelity with which $\hat A$-amplified squeezed vacuum (single-photon) state can approximate the even (odd) SCS of target amplitude $\alpha_f$ is 
\begin{align}
_{\rm max}F_{\pm S}^{\hat A}(\alpha_f)&=\mathop {\max }\limits_{r} {\big| {\langle { \pm_{\alpha_f} }
\big|
  \pm S_r^{\hat A} \rangle } \big|^2}.
\label{eq:maxFs}
\end{align}
The maximum fidelities $_{\rm max}F_{\pm S}^{\hat A}$ are 
numerically calculated \cite{numericalrecipes} and plotted in Fig.~\ref{fig:SqueezedFR} with respect to the desired amplitude $\alpha_f$ of the SCSs. 
Figures~5(a) and 5(c) show that higher fidelities to the even and odd SCSs  can be achieved using the amplification methods $\hat a \hat a^\dag$ and $\hat a^{\dag 2}$ on the squeezed vacuum and single-photon states, respectively, compared to the cases without the amplification methods.
Clearly, the  $\hat{a} \hat{a}^\dag$-amplification on the squeezed vacuum (single photon) state better approximates the even (odd) SCS of any amplitude with a higher fidelity than the squeezed vacuum (single photon) state itself does. 
While the $\hat a \hat a^\dag$ amplification is more efficient to 
approximate the  even and odd SCSs of small amplitude ($\alpha_f\lesssim1.81$ and $\alpha_f\lesssim2.32$, respectively), the ${\hat{a}^{\dag2}}$ amplification works better to increase the fidelities to  SCSs of larger amplitude as shown in Figs.~\ref{fig:SqueezedFR}(a) and \ref{fig:SqueezedFR}(c).

We also note that the required degrees of  squeezing, $r_{\pm S}^{\hat a \hat a ^\dag}  (\alpha_f)$ and $r_{\pm S}^{a ^{\dag2}}  (\alpha_f)$, of the ${\hat a \hat a ^\dag}$- and ${\hat a ^{\dag2}}$- amplified squeezed vacuum and single-photon states are significantly lessened as shown in Figs.~\ref{fig:SqueezedFR}(b) and \ref{fig:SqueezedFR}(d),
where $r=r_{\pm S}^{\hat A}  (\alpha_f)$ 
maximizes  fidelities $_{\rm max}F_{\pm S}^{\hat A}$ in Eq. (\ref{eq:maxFs}).
The amplification method $\hat a^{\dag2}$ requires less squeezing than ${\hat a \hat a ^\dag}$ does in order to approximate the even and odd SCSs.

\section{Remarks}
\label{Sec:Conclusion}

 We have suggested and investigated an efficient noiseless amplification scheme for coherent states with large amplitudes, while previous studies are suitable for amplification of coherent states with small amplitudes \cite{Ralph:2008wi, Fiurasek:2009jg, Xiang:2010hx, Zavatta:2011ea}.
Our scheme may be useful for various quantum information applications that require SCSs of sufficient large amplitudes.
As a specific example, this is closely related to efficient realization of fault-tolerant quantum computing \cite{Jeong:2002jj,Ralph:2003jk,Lund:2008hra}, where superpositions of coherent states (SCSs) with $\alpha\approx1.6$ (or $\alpha\approx2.26$ to generate entangled coherent states of amplitude $\approx1.6$) are required as optimized resources \cite{Lund:2008hra}. Our proposal enables one to efficiently amplify SCSs of $\alpha>0.755$ (for even SCSs) and $\alpha>1.324$ (for odd SCSs) in order to reach this size so that they become useful as optimized resources for quantum computing \cite{Jeong:2002jj,Ralph:2003jk,Lund:2008hra}.

The proposed amplification scheme utilizes two-photon addition, $\hat a^{\dag2}$, and its performance  was compared with that of a previous scheme using the photon addition and subtraction, $\hat a \hat a^\dag$, which is efficient  for amplifying small-amplitude coherent states \cite{Fiurasek:2009jg, Zavatta:2011ea}. 
We have found that $\hat a^{\dag2}$ is better than  $\hat a \hat a^\dag$, in terms of EIN,  as a noiseless amplifier for coherent states when the initial amplitudes are $\alpha_i\gtrsim0.91$. Similarly, when
$\hat a \hat a^\dag$ and  $\hat a^{\dag2}$ are successively applied with different possible orders,
$\hat a^{\dag2}$ is found to serve as the best noiseless amplifier as far as the initial amplitude is  $\alpha_i\gtrsim1.05$. Our study verifies that $\hat a^{\dag2}$ is also more noiseless for large SCSs than $\hat a \hat a^\dag$ in terms of optimal phase uncertainty. 
In addition, the ${\hat a \hat a ^\dag}$ and $\hat a^{\dag2}$ amplifications  can be applied to the squeezed vacuum and squeezed single-photon states in oder to approximate even and odd SCSs in a more efficient way. With the help of the amplification methods, SCSs with higher fidelities can be obtained using relatively small degrees of squeezing on the vacuum or single-photon states; the  $\hat a^{\dag2}$ amplification can be particularly useful to obtain SCSs of large amplitudes.

\appendix
\section*{Appendix A: Wigner functions of coherent states after photonic operations}
The Wigner function of a quantum state $\hat \rho$ is defined as \cite{Wigner1932}
\begin{align}
W(x, y) = \frac{2}{\pi} {\rm Tr}\Big[ \hat {D}^\dag (x+i y) \hat \rho  \hat {D} (x+i y) (-1)^{\hat n}\Big],
\end{align}
where $\hat {D}^\dag ( x+iy)$ is a displacement operator defined as
\begin{equation}
\hat {D}^\dag ( x+iy) = \exp [(x+i y)\hat a^\dag- (x-iy)\hat a]
\end{equation} 
The Wigner function of a coherent state with real amplitude $\alpha_i$ is calculated as
\begin{align}
W^{\hat 1} (x, y)= \frac{2 }{\pi}e^{-2 (x-\alpha_i)^2-2 y^2}.
\end{align}
After applying $\hat a^\dag$, the Wigner function becomes
\begin{align}
W^{\hat a^\dag} (x, y)= \frac{2 }{\pi}e^{-2 (x-\alpha_i)^2-2 y^2} \frac{({\alpha_i} -2 {x})^2+4 {y}^2-1}{1+\alpha_i^2}.
\label{eq:1PAWF}
\end{align}
When $\hat a$ is applied to Eq. (\ref{eq:1PAWF}), the Wigner function transforms into
\begin{align}
W^{\hat a \hat a^\dag} (x, y)= \frac{2 }{\pi}e^{-2 (x-\alpha_i)^2-2 y^2}
\frac{{\alpha_i} ^4+{\alpha_i} ^2 \left(4 {x}^2+4 {y}^2-3\right)-4 {\alpha_i} ^3 {x}+4 {\alpha_i}  {x}+1}{{\alpha_i} ^4+3 {\alpha_i} ^2+1}.
\end{align}
However when $\hat a^\dag$ is applied to Eq. (\ref{eq:1PAWF}) instead of $\hat a$, the Wigner function becomes 
\begin{align}
\nonumber
W^{\hat a \hat a^\dag} (x, y)&= \frac{2 }{\pi}e^{-2 (x-\alpha_i)^2-2 y^2}\\
&\times\frac{16 {{y}}^4+8 {y}^2 \left\{({\alpha_i} -2 {x})^2-2\right\}+({\alpha_i} -2 {x})^2 \left\{({\alpha_i} -2 {x})^2-4\right\}+2}{{\alpha_i} ^4+4 {\alpha_i} ^2+2}.
\end{align}

\section*{Appendix B: Fidelities, amplitude gains and EINs of amplified coherent states for two-cycle amplifications}
We have obtained the normalization factor in Eq.~(1) for the amplification of $\hat A \in \{(\hat{a} \hat{a}^\dag)^2, \hat{a}^{\dag4}, \hat{a} \hat{a}^\dag\hat{a}^{\dag2}, \hat{a}^{\dag2}\hat{a} \hat{a}^\dag\}$ with an initial coherent state of amplitude ${\alpha_i}$ as
\begin{equation}
\begin{split}
N^{ (\hat a \hat{a}^{\dag})^2}  ({\alpha_i})&=({\alpha_i}  ^8+10  {\alpha_i} ^6+25  {\alpha_i}^4+15  {\alpha_i} ^2+1)^{-\frac{1}{2}},
\\
N^{ \hat{a}^{\dag4}}  ({\alpha_i})&= ({\alpha_i}^8+16{\alpha_i} ^6+72  {\alpha_i} ^4+96 {\alpha_i} ^2+24)^{-\frac{1}{2}},
\\
N^{\hat a \hat{a}^{\dag}\hat{a}^{\dag2}}  ({\alpha_i})&= ( {\alpha_i} ^8+15{\alpha_i}^6+63 {\alpha_i} ^4+78{\alpha_i}  ^2+18)^{-\frac{1}{2}},
\\
N^{ \hat{a}^{\dag2}\hat a \hat{a}^{\dag}}  ({\alpha_i})&= ({\alpha_i}  ^8+11 {\alpha_i} ^6+31{\alpha_i} ^4+22{\alpha_i}^2+2)^{-\frac{1}{2}}.
\label{eq:twocycleNC}
\end{split}
\end{equation}
The fidelities
of the $\hat A$-amplified  coherent state
  to the coherent state of amplitude $\alpha_f$  are then calculated as
\begin{equation}
\begin{split}
F^{(\hat a \hat a^\dag)^2}&= {\left\{N^{ (\hat a \hat{a}^{\dag})^2}  ({\alpha_i})\right\}^2}{e^{-(  {\alpha_i} -\alpha _f)^2} ( {\alpha_i} ^2 \alpha _f^2+3  {\alpha_i}  \alpha _f+1)^2},\\
F^{ \hat a^{\dag4}}&= {\left\{N^{ \hat{a}^{\dag4}}  ({\alpha_i})\right\}^2}{ {\alpha_i}  ^8 e^{-({\alpha_i}  - \alpha _f){}^2}},\\
F^{\hat a \hat{a}^{\dag}\hat{a}^{\dag2}}  &= {\left\{ N^{\hat a \hat{a}^{\dag}\hat{a}^{\dag2}}  ({\alpha_i}) \right\}^2}{\alpha _f^4 e^{-( {\alpha_i}  -\alpha _f )^2}  ({\alpha_i} \alpha _f+3)^2}, \\
F^{\hat a^{\dag2}\hat a \hat a^\dag}&={\left\{ N^{ \hat{a}^{\dag2}\hat a \hat{a}^{\dag}}  ({\alpha_i}) \right\}^2}{ \alpha _f ^4 e^{-( {\alpha_i}  - \alpha _f )^2} ( {\alpha_i}   \alpha _f +1)^2}.
\label{eq:twocycleFC}
\end{split}
\end{equation}
The gains from the amplification $\hat A$ are obtained using  Eq. (8) as
\begin{equation}
\begin{split}
g^{ (\hat a \hat{a}^{\dag})^2}_\lambda &={\left\{N^{ (\hat a \hat{a}^{\dag})^2}  ({\alpha_i})\right\}^2} (\alpha_i ^8+12 \alpha_i ^6+38 \alpha_i ^4+32 \alpha_i ^2+4),
\\
g^{ \hat{a}^{\dag4}}_\lambda &={ {\left\{N^{ \hat{a}^{\dag4}}  ({\alpha_i})\right\}^2}}(\alpha_i ^8+20 \alpha_i ^6+120 \alpha_i ^4+240 \alpha_i ^2+120),
\\
g^{\hat a \hat{a}^{\dag}\hat{a}^{\dag2}}_\lambda&= { {\left\{N^{\hat a \hat{a}^{\dag}\hat{a}^{\dag2}}  ({\alpha_i})\right\}^2}} (\alpha_i ^8+18 \alpha_i ^6+96 \alpha_i ^4+168 \alpha_i ^2+72),
\\
g^{ \hat{a}^{\dag2}\hat a \hat{a}^{\dag}}_\lambda &={\left\{N^{\hat{a}^{\dag2}\hat a \hat{a}^{\dag}}  ({\alpha_i})\right\}^2}(\alpha_i ^8+14 \alpha_i ^6+54 \alpha_i ^4+60 \alpha_i ^2+12),
\label{eq:twocyclegainCA}
\end{split}
\end{equation}
and the EINs are obtained using Eq. (11) as
\begin{equation}
\begin{split}
E^{ (\hat a \hat{a}^{\dag})^2}_\lambda  ({\alpha_i})&=-\Big\{4 {\alpha_i}^{14}+62 {\alpha_i}^{12}+382 {\alpha_i}^{10}+1101 {\alpha_i}^8+1554 {\alpha_i}^6+955 {\alpha_i}^4+226 {\alpha_i}^2\\
&+2 \left(2 {\alpha_i}^{12}+26 {\alpha_i}^{10}+125 {\alpha_i}^8+254 {\alpha_i}^6+225 {\alpha_i}^4+66 {\alpha_i}^2+7\right) {\alpha_i}^2 \cos (2 \lambda )+15\Big\}\\
&/\Big\{2 \left({\alpha_i}^8+12 {\alpha_i}^6+38 {\alpha_i}^4+32 {\alpha_i}^2+4\right)^2\Big\},
\\
E^{ \hat{a}^{\dag4}}_\lambda  ({\alpha_i})&=-4\Big\{ {\alpha_i}^{14}+26 {\alpha_i}^{12}+276 {\alpha_i}^{10}+1488 {\alpha_i}^8+4344 {\alpha_i}^6+6624 {\alpha_i}^4+4896 {\alpha_i}^2\\&+\left({\alpha_i}^{12}+24 {\alpha_i}^{10}+228 {\alpha_i}^8+1056 {\alpha_i}^6+2520 {\alpha_i}^4+2880 {\alpha_i}^2+1440\right) {\alpha_i}^2 \cos (2 \lambda )+1152\Big\}\\
&/\Big\{\left({\alpha_i}^8+20 {\alpha_i}^6+120 {\alpha_i}^4+240 {\alpha_i}^2+120\right)^2\Big\},
\end{split}
\label{eq:twocycleEINC}
\end{equation}
\begin{equation}
\begin{split}
E^{\hat a \hat{a}^{\dag}\hat{a}^{\dag2}}_\lambda  ({\alpha_i})&= -3\Big\{2 {\alpha_i}^{14}+49 {\alpha_i}^{12}+486 {\alpha_i}^{10}+2421 {\alpha_i}^8+6432 {\alpha_i}^6+8784 {\alpha_i}^4+5688 {\alpha_i}^2\\
&+2 \left({\alpha_i}^{12}+22 {\alpha_i}^{10}+189 {\alpha_i}^8+780 {\alpha_i}^6+1626 {\alpha_i}^4+1584 {\alpha_i}^2+648\right) {\alpha_i}^2 \cos (2 \lambda )+1188\Big\}\\&/\Big\{2 \left({\alpha_i}^8+18 {\alpha_i}^6+96 {\alpha_i}^4+168 {\alpha_i}^2+72\right)^2\Big\},
\\
E^{ \hat{a}^{\dag2}\hat a \hat{a}^{\dag}}_\lambda  ({\alpha_i})&=-\Big\{6 {\alpha_i}^{14}+103 {\alpha_i}^{12}+714 {\alpha_i}^{10}+2395 {\alpha_i}^8+4128 {\alpha_i}^6+3344 {\alpha_i}^4+1192 {\alpha_i}^2\\
&+2 \left(3 {\alpha_i}^{12}+46 {\alpha_i}^{10}+271 {\alpha_i}^8+716 {\alpha_i}^6+886 {\alpha_i}^4+400 {\alpha_i}^2+72\right) {\alpha_i}^2 \cos (2 \lambda )+124\Big\}\\
&/\Big\{2 \left({\alpha_i}^8+14 {\alpha_i}^6+54 {\alpha_i}^4+60 {\alpha_i}^2+12\right)^2\Big\}.
\end{split}
\label{eq:twocycleEINC2}
\end{equation}


\section*{Appendix C: Quantum Fisher information of amplified SCSs}

The values of the quantum Fisher information after applying $\hat A \in \{\hat 1, \hat{a} \hat{a}^\dag, \hat{a}^{\dag2}\}$ to SCSs of amplitude $\alpha_i$ are
\begin{equation*}
\begin{split}
	{\cal{F}}_+^{\hat 1} (\alpha_i)=&\frac{4 {\alpha_i} ^2 \left(4 e^{2 {\alpha_i} ^2} {\alpha_i} ^2+e^{4 {\alpha_i} ^2}-1\right)}{\left(e^{2 {\alpha_i} ^2}+1\right)^2},
\\
	{\cal{F}}_+^{\hat a\hat{a}^\dag} (\alpha_i)=&16{(N_+^{\hat a\hat{a}^\dag})^4} e^{-2 {\alpha_i} ^2} {\alpha_i} ^2 \Big[ (N_+^{\hat a\hat{a}^\dag})^{-2} e^{\alpha_i ^2} \left\{4 \left(2 \alpha_i ^4+1\right) \sinh \alpha_i ^2+\left(\alpha_i ^4+14\right) \alpha_i ^2 \cosh \alpha_i ^2\right\} \\
	&-4 \left\{\left({\alpha_i} ^4+4\right) {\alpha_i}  \sinh {\alpha_i} ^2 +5 {\alpha_i} ^3 \cosh {\alpha_i} ^2 \right\}^2\Big],
\\
	{\cal{F}}_+^{\hat{a}^{\dag2}} (\alpha_i)=&16 {({N_+^{\hat{a}^{\dag 2}}})^4} e^{-2 {\alpha_i} ^2} \Big[e^{{\alpha_i} ^2} ({N_+^{\hat{a}^{\dag 2}}})^{-2} \left\{\left(13 {\alpha_i} ^4+46\right) {\alpha_i} ^2 \sinh {\alpha_i} ^2 +\left({\alpha_i} ^8+46 {\alpha_i} ^4+8\right) \cosh {\alpha_i} ^2 \right\}\\
	&-4 \left\{\left({\alpha_i} ^4+14\right) {\alpha_i} ^2 \sinh {\alpha_i} ^2+ 4 \left(2 {\alpha_i} ^4+1\right) \cosh {\alpha_i} ^2 \right\}^2\Big]
	\label{eq:FisherA}
\end{split}
\end{equation*}
\begin{equation*}
\begin{split}
		{\cal{F}}_-^{\hat 1} (\alpha_i)=& \frac{4 {\alpha_i} ^2 \left(-4 e^{2 {\alpha_i} ^2} {\alpha_i} ^2+e^{4 {\alpha_i} ^2}-1\right)}{\left(e^{2 {\alpha_i} ^2}-1\right)^2},
\\
	{\cal{F}}_-^{\hat a\hat{a}^\dag} (\alpha_i)=&16 {({N_-^{\hat a \hat{a}^\dag}})^4} e^{-2 {\alpha_i} ^2} {\alpha_i} ^2 \Big[ ({N_-^{\hat a \hat{a}^\dag}})^{-2} e^{{\alpha_i} ^2}
	\left\{\left(\alpha_i ^4+14\right) \alpha_i ^2 \sinh \alpha_i ^2+4\left(2 \alpha_i ^4+1\right) \cosh \alpha_i ^2\right\}
	\\
	&-4 \left(\left({\alpha_i} ^4+4\right) {\alpha_i}  \cosh {\alpha_i} ^2+5 {\alpha_i} ^3 \sinh {\alpha_i} ^2 \right)^2\Big],
\\
	{\cal{F}}_-^{\hat{a}^{\dag2}} (\alpha_i)=&16 {({N_-^{\hat{a}^{\dag 2}}})^4} e^{-2 {\alpha_i} ^2} \Big[e^{{\alpha_i} ^2} ({N_-^{\hat{a}^{\dag 2}}})^{-2} \left\{\left(13 {\alpha_i} ^4+46\right) {\alpha_i} ^2 \cosh {\alpha_i} ^2+\left({\alpha_i} ^8+46 {\alpha_i} ^4+8\right) \sinh {\alpha_i} ^2 \right\}\\
	&-4 \left\{4 \left(2 {\alpha_i} ^4+1\right) \sinh {\alpha_i} ^2+\left({\alpha_i} ^4+14\right) {\alpha_i} ^2 \cosh {\alpha_i} ^2 \right\}^2\Big],
\label{eq:FisherB}
\end{split}
\end{equation*}
where subscripts $+$ and $-$ correspond to the even and odd SCSs, respectively.

\section*{Acknowledgments}
The authors thank  H. Kwon, C. Park, S. Yang, Y. Lim, D. Oi and T. Spiller for useful discussions. 
This work was supported by the National Research Foundation of Korea (NRF) grant funded by the Korea government (MSIP) (No. 2010-0018295)
and by the EU under the ERA-NET CHIST-ERA project QSCALE.


\begin{thebibliography}{10}
\newcommand{\enquote}[1]{``#1''}

\bibitem{Wootters:1982ex}
W.~K. Wootters and W.~H. Zurek, \enquote{{A single quantum cannot be cloned},}
  Nature Photon. \textbf{299}, 802--803 (1982).

\bibitem{Herbert:1982br}
N.~Herbert, \enquote{{FLASH{\textemdash}A superluminal communicator based upon
  a new kind of quantum measurement},} Found. Phys. \textbf{12}, 1171--1179
  (1982).

\bibitem{Arthurs:1988fv}
E.~Arthurs and M.~Goodman, \enquote{{Quantum Correlations: A Generalized
  Heisenberg Uncertainty Relation},} Phys. Rev. Lett. \textbf{60}, 2447--2449
  (1988).

\bibitem{Caves:1982hd}
C.~Caves, \enquote{{Quantum limits on noise in linear amplifiers},} Phys. Rev.
  D \textbf{26}, 1817--1839 (1982).

\bibitem{Ralph:2008wi}
T.~C. Ralph and A.~P. Lund, \enquote{{Nondeterministic Noiseless Linear
  Amplification of Quantum Systems},} in {Proceedings of 9th
  International Conference on Quantum Communication, Measurement, and
  Computing,} A.~Lvovsky, ed. (AIP, 2009), pp. 155--160.

\bibitem{Fiurasek:2009jg}
J.~Fiur{\'a}{\v s}ek, \enquote{{Engineering quantum operations on traveling
  light beams by multiple photon addition and subtraction},} Phys. Rev. A
  \textbf{80}, 053822 (2009).
  
\bibitem{Ferreyrol:2010}
F.~Ferreyrol, M.~Barbieri, R.~Blandino, S.~Fossier, R.~Tualle-Brouri, and P.~Grangier,
  \enquote{{Implementation of a Nondeterministic Optical Noiseless Amplifier},} Phys. Rev. Lett. \textbf{104}, 123603 (2010).

\bibitem{Xiang:2010hx}
G.~Y. Xiang, T.~C. Ralph, A.~P. Lund, N.~Walk, and G.~J. Pryde,
  \enquote{{Heralded noiseless linear amplification and distillation of
  entanglement},} Nature Photon. \textbf{4}, 316--319 (2010).

\bibitem{Zavatta:2011ea}
A.~Zavatta, J.~Fiur{\'a}{\v s}ek, and M.~Bellini, \enquote{{A high-fidelity
  noiseless amplifier for quantum light states},} Nature Photon. \textbf{5},
  52--60 (2011).

\bibitem{Fossier:2009dz}
S.~Fossier, E.~Diamanti, T.~Debuisschert, R.~Tualle-Brouri, and P.~Grangier,
  \enquote{{Improvement of continuous-variable quantum key distribution systems
  by using optical preamplifiers},} J. Phys. B: At. Mol. Opt. Phys.
  \textbf{42}, 114014 (2009).

\bibitem{Blandino:2012bg}
R.~Blandino, A.~Leverrier, M.~Barbieri, J.~Etesse, P.~Grangier, and
  R.~Tualle-Brouri, \enquote{{Improving the maximum transmission distance of
  continuous-variable quantum key distribution using a noiseless amplifier},}
  Phys. Rev. A \textbf{86}, 012327 (2012).

\bibitem{Micuda:2012em}
M.~Mi{\v c}uda, I.~Straka, M.~Mikov{\'a}, M.~Du{\v s}ek, N.~J. Cerf,
  J.~Fiur{\'a}{\v s}ek, and M.~Je{\v z}ek, \enquote{{Noiseless Loss Suppression
  in Quantum Optical Communication},} Phys. Rev. Lett. \textbf{109}, 180503
  (2012).

\bibitem{Duan:2001dt}
L.~M. Duan, M.~D. Lukin, J.~I. Cirac, and P.~Zoller, \enquote{{Long-distance
  quantum communication with atomic ensembles and linear optics},} Nature
  Photon. \textbf{414}, 413--418 (2001).

\bibitem{Usuga:2010fw}
M.~A. Usuga, C.~R. M{\"u}ller, C.~Wittmann, P.~Marek, R.~Filip, C.~Marquardt,
  G.~Leuchs, and U.~L. Andersen, \enquote{{Noise-powered probabilistic
  concentration of phase information},} Nature Phys. \textbf{6}, 767--771
  (2010).

\bibitem{Ralph:2011ct}
T.~C. Ralph, \enquote{{Quantum error correction of continuous-variable states
  against Gaussian noise},} Phys. Rev. A \textbf{84}, 022339 (2011).

\bibitem{Leonhardt:1994fm}
U.~Leonhardt and H.~Paul, \enquote{{High-Accuracy Optical Homodyne Detection
  with Low-Efficiency Detectors: "Preamplification" from Antisqueezing},} Phys.
  Rev. Lett. \textbf{72}, 4086--4089 (1994).

\bibitem{Park:2012eq}
J.~Park, S.-Y. Lee, H.-W. Lee, and H.~Nha, \enquote{{Enhanced Bell violation by
  a coherent superposition of photon subtraction and addition},} J. Opt. Soc.
  Am. B \textbf{29}, 906--911 (2012).

\bibitem{Torlai:2013hw}
G.~Torlai, G.~McKeown, P.~Marek, R.~Filip, H. Jeong, M. Paternostro, and G. D. Chiara,
  \enquote{{Violation of Bell's inequalities with preamplified homodyne
  detection},} Phys. Rev. A  \textbf{87}, 052112 (2013).

\bibitem{Pegg:1998}
D.~T.~Pegg, L.~S.~Phillips, and S.~M.~Barnett, \enquote{{Optical state truncation by projection
synthesis},} Phys. Rev. Lett. \textbf{81}, 1604--1606 (1998).

\bibitem{Wenger:2004cw}
J.~Wenger, R.~Tualle-Brouri, and P.~Grangier, \enquote{{Non-Gaussian Statistics
  from Individual Pulses of Squeezed Light},} Phys. Rev. Lett. \textbf{92},
  153601 (2004).



\bibitem{Zavatta:2004km}
A.~Zavatta, S.~Viciani, and M.~Bellini, \enquote{{Quantum-to-classical
  transition with single-photon-added coherent states of light.}} Science
  \textbf{306}, 660--662 (2004).

\bibitem{Mann:1995jy}
A.~Mann, B.~C. Sanders, and W.~J. Munro, \enquote{{Bell{\textquoteright}s
  inequality for an entanglement of nonorthogonal states},} Phys. Rev. A
  \textbf{51}, 989--991 (1995).

\bibitem{Filip:2001gw}
R.~Filip, J.~Reh{\'a}cek, and M.~Dusek, \enquote{{Entanglement of coherent
  states and decoherence},} J. Opt. B: Quantum Semiclass. Opt. \textbf{3},
  341--345 (2001).

\bibitem{Wilson:2002fl}
D.~Wilson, H.~Jeong, and M.~S. Kim, \enquote{{Quantum nonlocality for a mixed
  entangled coherent state},} J. Mod. Opt. \textbf{49}, 851--864
  (2002).

\bibitem{Wenger:2003eu}
J.~Wenger, M.~Hafezi, F.~Grosshans, R.~Tualle-Brouri, and P.~Grangier,
  \enquote{{Maximal violation of Bell inequalities using continuous-variable
  measurements},} Phys. Rev. A \textbf{67}, 012105 (2003).

\bibitem{Jeong:2003es}
H.~Jeong, W.~Son, and M.~S. Kim, \enquote{{Quantum nonlocality test for
  continuous-variable states with dichotomic observables},} Phys. Rev. A
  \textbf{67}, 012106 (2003).

\bibitem{Stobinska:2007co}
H.~Jeong and T.~C. Ralph, \enquote{{Violation of Bell{\textquoteright}s
  inequality using classical measurements and nonlinear local operations},}
  Phys. Rev. A \textbf{75}, 052105 (2007).

\bibitem{Jeong:2008jy}
H.~Jeong, \enquote{{Testing Bell inequalities with photon-subtracted Gaussian
  states},} Phys. Rev. A \textbf{78}, 042101 (2008).

\bibitem{Lee:2009kt}
C.-W. Lee and H.~Jeong, \enquote{{Effects of squeezing on quantum nonlocality
  of superpositions of coherent states},} Phys. Rev. A \textbf{80}, 052105
  (2009).

\bibitem{Gerry:2009gp}
C.~C. Gerry, J.~Mimih, and A.~Benmoussa, \enquote{{Maximally entangled coherent
  states and strong violations of Bell-type inequalities},} Phys. Rev. A
  \textbf{80}, 022111 (2009).

\bibitem{Jeong:2009dl}
H.~Jeong and T.~C. Ralph, \enquote{{Failure of Local Realism Revealed by
  Extremely-Coarse-Grained Measurements},} Phys. Rev. Lett. \textbf{102},
  060403 (2009).

\bibitem{Paternostro:2010ca}
M.~Paternostro and H.~Jeong, \enquote{{Testing nonlocal realism with entangled
  coherent states},} Phys. Rev. A \textbf{81}, 032115 (2010).

\bibitem{Lee:2011kx}
C.-W. Lee and H.~Jeong, \enquote{{Faithful test of nonlocal realism with
  entangled coherent states},} Phys. Rev. A \textbf{83}, 022102 (2011).

\bibitem{Lim:2012kj}
Y.~Lim, M.~Kang, J.~Lee, and H.~Jeong, \enquote{{Using macroscopic entanglement
  to close the detection loophole in Bell-inequality tests},} Phys. Rev. A
  \textbf{85}, 062112 (2012).

\bibitem{Kirby:2013ev}
B.~T. Kirby and J.~D. Franson, \enquote{{Nonlocal interferometry using
  macroscopic coherent states and weak nonlinearities},} Phys. Rev. A
  \textbf{87}, 053822 (2013).

\bibitem{Kirby:2014bd}
B.~T. Kirby and J.~D. Franson, \enquote{{Macroscopic state interferometry over
  large distances using state discrimination},} Phys. Rev. A \textbf{89},
  033861 (2014).

\bibitem{Park:2015kl}
C.-Y. Park and H.~Jeong, \enquote{{Bell-inequality tests using asymmetric
  entangled coherent states in asymmetric lossy environments},} Phys. Rev. A
  \textbf{91}, 042328 (2015).

\bibitem{Gerry:2001ko}
C.~Gerry and R.~Campos, \enquote{{Generation of maximally entangled photonic
  states with a quantum-optical Fredkin gate},} Phys. Rev. A \textbf{64},
  063814 (2001).

\bibitem{Gerry:2002hm}
C.~Gerry, A.~Benmoussa, and R.~Campos, \enquote{{Nonlinear interferometer as a
  resource for maximally entangled photonic states: Application to
  interferometry},} Phys. Rev. A \textbf{66}, 013804 (2002).

\bibitem{Ralph:2002ip}
T.~C. Ralph, \enquote{{Coherent superposition states as quantum rulers},} Phys.
  Rev. A \textbf{65}, 042313 (2002).

\bibitem{Munro:2002ew}
W.~J. Munro, K.~Nemoto, G.~J. Milburn, and S.~Braunstein, \enquote{{Weak-force
  detection with superposed coherent states},} Phys. Rev. A \textbf{66}, 023819
  (2002).

\bibitem{Campos:2003ka}
R.~Campos, C.~Gerry, and A.~Benmoussa, \enquote{{Optical interferometry at the
  Heisenberg limit with twin Fock states and parity measurements},} Phys. Rev.
  A \textbf{68}, 023810 (2003).

\bibitem{Joo:2011ge}
J.~Joo, W.~J. Munro, and T.~P. Spiller, \enquote{{Quantum Metrology with
  Entangled Coherent States},} Phys. Rev. Lett. \textbf{107}, 083601 (2011).

\bibitem{Hirota:2011vl}
O.~Hirota, K.~Kato, and D.~Murakami, \enquote{{Effectiveness of entangled
  coherent state in quantum metrology},}  http://arxiv.org/abs/1108.1517 (2011).

\bibitem{Joo:2012fh}
J.~Joo, K.~Park, H.~Jeong, W.~J. Munro, K.~Nemoto, and T.~P. Spiller,
  \enquote{{Quantum metrology for nonlinear phase shifts with entangled
  coherent states},} Phys. Rev. A \textbf{86}, 043828 (2012).

\bibitem{Zhang:2013jk}
Y.~M. Zhang, X.~W. Li, W.~Yang, and G.~R. Jin, \enquote{{Quantum Fisher
  information of entangled coherent states in the presence of photon loss},}
  Phys. Rev. A \textbf{88}, 043832 (2013).

\bibitem{Cochrane:1999ck}
P.~T. Cochrane, G.~J. Milburn, and W.~J. Munro, \enquote{{Macroscopically
  distinct quantum-superposition states as a bosonic code for amplitude
  damping},} Phys. Rev. A \textbf{59}, 2631--2634 (1999).

\bibitem{Jeong:2002jj}
H.~Jeong and M.~S. Kim, \enquote{{Efficient quantum computation using coherent
  states},} Phys. Rev. A \textbf{65}, 042305 (2002).

\bibitem{Ralph:2003jk}
T.~C. Ralph, A.~Gilchrist, G.~J. Milburn, W.~J. Munro, and S.~Glancy,
  \enquote{{Quantum computation with optical coherent states},} Phys. Rev. A
  \textbf{68}, 042319 (2003).

\bibitem{Lund:2008hra}
A.~P. Lund, T.~C. Ralph, and H.~L. Haselgrove, \enquote{{Fault-Tolerant Linear
  Optical Quantum Computing with Small-Amplitude Coherent States},} Phys. Rev.
  Lett. \textbf{100}, 030503 (2008).


\bibitem{Myers:2011ef}
C.~R. Myers and T.~C. Ralph, \enquote{{Coherent state topological cluster state
  production},} New J. Phys. \textbf{13}, 115015 (2011).

\bibitem{Kim:2015go}
J.~Kim, J.~Lee, S.-W. Ji, H.~Nha, P.~M. Anisimov, and J.~P. Dowling,
  \enquote{{Coherent-state optical qudit cluster state generation and
  teleportation via homodyne detection},} Opt. Commun. \textbf{337},
  79--82 (2015).


\bibitem{vanEnk:2001gb}
S.~J. van Enk and O.~Hirota, \enquote{{Entangled coherent states: Teleportation
  and decoherence},} Phys. Rev. A \textbf{64}, 022313 (2001).

\bibitem{Jeong:2001eu}
H.~Jeong, M.~S. Kim, and J.~Lee, \enquote{{Quantum-information processing for a
  coherent superposition state via a mixedentangled coherent channel},} Phys.
  Rev. A  \textbf{64}, 052308(2001).



\bibitem{Simon:2014ih}
D.~S. Simon, G.~Jaeger, and A.~V. Sergienko,
  \enquote{{Entangled-coherent-state quantum key distribution with entanglement
  witnessing},} Phys. Rev. A \textbf{89}, 012315 (2014).

\bibitem{Lund:2004tg}
A.~P. Lund, H.~Jeong, T.~C. Ralph, and M.~S. Kim, \enquote{{Conditional
  production of superpositions of coherent states with inefficient photon
  detection},} Phys. Rev. A \textbf{70}, 020101 (2004).

\bibitem{Jeong:2005kq}
H.~Jeong, A.~P. Lund, and T.~C. Ralph, \enquote{{Production of superpositions
  of coherent states in traveling optical fields with inefficient photon
  detection},} Phys. Rev. A \textbf{72}, 013801 (2005).

\bibitem{Ourjoumtsev:2006ima}
A.~Ourjoumtsev, R.~Tualle-Brouri, J.~Laurat, and P.~Grangier,
  \enquote{{Generating optical Schr{\"o}dinger kittens for quantum information
  processing.}} Science \textbf{312}, 83--86 (2006).

\bibitem{Ourjoumtsev:2009jh}
A.~Ourjoumtsev, F.~Ferreyrol, R.~Tualle-Brouri, and P.~Grangier, \enquote{{Preparation of non-local superpositions of quasi-classical light states},} Nature Phys. \textbf{5}, 189--192
  (2009).


\bibitem{Joo:2015wn}
J.~Joo, M.~Elliott, D.~K.~L. Oi, E.~Ginossar, and T.~P. Spiller,
  \enquote{{Deterministic amplification of Schr{\"o}dinger cat states in
  circuit quantum electrodynamics},}  http://arxiv.org/abs/1502.06782 (2015).

\bibitem{Grangier:1992fva}
P.~Grangier, J.-M. Courty, and S.~Reynaud, \enquote{{Characterization of
  nonideal quantum non-demolition measurements},} Opt. Commun.
  \textbf{89}, 99--106 (1992).

\bibitem{numericalrecipes}
W. H. Press, B. P. Flannery, S. A. Teukolsky, and W. T. Vetterling, {\it Numerical Recipes} (Cambridge University, 1988).

\bibitem{Bellini:2010}
M.~ Bellini, and A. Zavatta, \enquote{{Manipulating Light States by Single-Photon Addition and Subtraction},} Prog. Optics
  \textbf{55}, 41--83 (2010).



\bibitem{Parigi:2007}
V.~Parigi, A.~Zavatta, M.~Kim, and M.~Bellini \enquote{{Probing Quantum Commutation Rules by Addition and Subtraction of Single Photons to/from a Light Field},} Science
  \textbf{317}, 1890 (2007).
  

\bibitem{Braunstein:1994ug}
S.~L. Braunstein and C.~M. Caves, \enquote{{Statistical distance and the
  geometry of quantum states},} Phys. Rev. Lett. \textbf{72}, 3439--3443
  (1994).

\bibitem{Braunstein:1996ke}
S.~L. Braunstein, C.~M. Caves, and G.~J. Milburn, \enquote{{Generalized
  Uncertainty Relations: Theory, Examples, and Lorentz Invariance},} Ann. Phys. \textbf{247}, 135--173 (1996).
    

\bibitem{Marek08} P. Marek, H. Jeong, M.~S. Kim,  \enquote{{Generating ``squeezed'' superpositions of coherent states using photon addition and subtraction},} Phys. Rev. A \textbf{78}, 063811 (2008).

\bibitem{Wigner1932}
E.~Wigner, \enquote{{On the quantum correction for thermodynamic equilibrium},} Phys. Rev. \textbf{40}, 749
  (1932).


\bibitem{nf01} S. R. Huisman, N. Jain, S. A. Babichev, Frank Vewinger, A. N. Zhang, S. H. Youn, and A. I. Lvovsky,
\enquote{{ Instant single-photon Fock state tomography},}
   Optics Letters \textbf{34}, 2739 (2009).

\bibitem{nf02} M. Cooper, L. J. Wright, C. Soller, and B.J. Smith, 
\enquote{{ 
Experimental generation of multiphoton fock states},}
Optics Express \textbf{21}, 5309 (2012).


\end{thebibliography}
\end{document}